*In situ characterization of vacancy ordering in Ge-Sb-Te phase-change memory alloys*

Ting-Ting Jiang[a#], Xu-Dong Wang[a#], Jiang-Jing Wang[a,b], Han-Yi Zhang[a], Lu Lu[c], Chunlin Jia[c], Matthias Wuttig[b,d,e], Riccardo Mazzarello[f], Wei Zhang[a]*, En Ma[a]

[a] Center for Alloy Innovation and Design (CAID), State Key Laboratory for Mechanical Behavior of Materials, Xi'an Jiaotong University, Xi'an, 710049, China.
[b] Institute of Physics IA, RWTH Aachen University, Aachen, 52074, Germany
[c] The School of Microelectronics, State Key Laboratory for Mechanical Behavior of Materials, Xi'an Jiaotong, University, Xi'an, 710049, China
[d] JARA-FIT and JARA-HPC, RWTH Aachen University, Aachen, 52056, Germany
[e] Peter Grünberg Institute (PGI 10), Forschungszentrum Jülich GmbH, Jülich, 52425, Germany
[f] Department of Physics, Sapienza University of Rome, Rome, 00185, Italy

[#]These authors contributed equally to this work.
*E-mail: wzhang0@mail.xjtu.edu.cn

**Abstract:**
Tailoring the degree of structural disorder in Ge-Sb-Te alloys is important for the development of non-volatile phase-change memory and neuro-inspired computing. Upon crystallization from the amorphous phase, these alloys form a cubic rocksalt-like structure with a high content of intrinsic vacancies. Further thermal annealing results in a gradual structural transition towards a layered structure and an insulator-to-metal transition. In this work, we elucidate the atomic-level details of the structural transition in crystalline $GeSb_2Te_4$ by *in situ* high-resolution transmission electron microscopy (HRTEM) experiments and *ab initio* density functional theory (DFT) calculations, providing a comprehensive real-time and real-space view of the vacancy ordering process. We also discuss the impact of vacancy ordering on altering the electronic and optical properties of $GeSb_2Te_4$, which is relevant to multilevel storage applications. The phase evolution paths in Ge-Sb-Te alloys and $Sb_2Te_3$ are illustrated using a summary diagram, which serves as a guide for designing phase-change memory devices.

1. **Introduction**

Chalcogenide phase-change materials (PCMs) are widely exploited for non-volatile memory (NVM) and in-memory computing (IMC) applications [1-10]. The basic principle is to utilize their capability of rapid phase transitions between amorphous and crystalline states, as well as the associated change in electrical resistance or optical reflectivity, to encode information. Ge-Sb-Te alloys along the GeTe-$Sb_2Te_3$ pseudo-binary line are the most widely investigated and used PCMs [11]. The atomic structure and defects of Ge-Sb-Te alloys have been extensively studied by X-ray diffraction (XRD) [11-13], transmission electron microscopy (TEM) experiments [14-20] and density functional theory (DFT) calculations [21-27]. Upon rapid crystallization of the amorphous (amor-) phase in devices [14], Ge-Sb-Te alloys form a metastable cubic rocksalt-like (cub-) crystalline phase with the anion-like sublattice occupied by Te atoms and the cation-like sublattice occupied by Ge, Sb and atomic vacancies. Further thermal annealing drives a gradual phase transformation from the cub-phase towards a stable trigonal phase (also termed as hexagonal, hex-, in literature) with alternately stacked Ge-Sb-Te blocks and van der Waals (vdW)-like gaps.

An unconventionally high amount of atomic vacancies, e.g. 12.5% in $GeSb_2Te_4$ (abbreviated as "GST" in the following), is found in the cub-phase. These vacancies reduce the unfavorable antibonding interactions that would be present for a non-stoichiometric phase with fully occupied Ge/Sb sublattice [28]. The resulting electronic configuration yields on average 3 $p$ electrons per site in the cubic rocksalt lattice [29]. Since the atomic arrangement is octahedral-like, the σ–bonds formed between adjacent atoms are only occupied by one electron (half an electron pair). This unconventional bonding configuration differs significantly from ordinary covalent bonding. It is better described as a distinct bonding mechanism, recently coined as metavalent bonding (MVB), which is characterized by the competition between electron delocalization as in metallic bonding and electron localization as in ionic and covalent bonding [30-34]. Vacancies can have a profound impact on this competition. The presence of atomic vacancies was unambiguously demonstrated by element-resolved atomic imaging experiments [18]. The statistical distribution of the intrinsic atomic vacancies induces Anderson-like localization of electrons at the Fermi level. Ordering of these vacancies drives a cub-to-hex structural transition and an insulator-to-metal transition[35-37].

The vacancy ordering process was first predicted on the basis of DFT calculations [36], and then observed in *ex situ* TEM experiments upon thermal annealing [18-20]. *In situ* heating TEM experiments were also reported in the literature [38-41]. However, the microscopic details of the vacancy migration and ordering are still lacking. In this work, we carry out *in situ* high-resolution TEM experiments in combination with electron beam irradiation, which enable a direct observation of the structural transition process locally. Together with insight provided via DFT calculations, our experimental analysis offers a comprehensive atomic picture of the cub-to-hex structural transitions in GST.

## 2. Experimental and simulation methods

The amor-$GeSb_2Te_4$ thin films of ~80 nm thickness were prepared on carbon-supported TEM grids via magnetron sputtering deposition from stoichiometric targets under 20 W power with a base pressure of $2\times10^{-6}$ mbar and an argon gas flow of 20 sccm. Then the films were covered by an electron-transparent $ZnS$-$SiO_2$ layer. The as-deposited films were annealed in a tube furnace under constant argon-flow of 1200 sccm at 150 °C for 1 hour, forming the cub-phase. The *in situ* electron beam irradiation experiments were performed for the cubic samples in a JEOL JEM-2100F TEM operated under 200 keV and in a JEM-200 CX TEM under 120 keV. For *ex situ* experiments, an amor-$GeSb_2Te_4$ thin film of ~450 nm-thick was deposited with magnetron sputtering on a silicon substrate, and was annealed in a tube furnace with an argon flow rate of 1200 sccm at different temperatures, namely, 175, 225, 250 and 350 °C, respectively, for 1 hour. After each annealing step, the thin film was used to prepare TEM sample specimens (~80 nm thickness) using a Helios Nano 600 dual-beam FIB system (FEI, Hillsboro, OR, USA) under 30 keV with beam current of Ga ions sequentially decreasing from 6.5 nA for coarse cutting to 28 pA for fine thinning. The energy dispersive X-ray (EDX) spectroscopy analysis confirms no Ga signal in the sample. The microstructures of all samples were characterized by bright-field and high-resolution TEM (HRTEM) imaging, selected area electron diffraction (SAED) analysis and fast Fourier transform (FFT) of HRTEM images in JEOL JEM-2100F TEM. Aberration corrected high-angle annular dark-field scanning transmission electron microscopy (HAADF-STEM) experiments were conducted at 200 keV on an JEOL-ARM200F TEM equipped with an EDX system.

DFT calculations were carried out for $GeSb_2Te_4$ models of different size. Large models with 1008

atoms were built for electron localization calculations, which were performed using the CP2K package [42]. Goedecker pseudopotentials [43] and the Perdew–Burke–Ernzerhof (PBE) functional [44] were used, and the Brillouin zone was sampled at the Γ point. A triple-zeta plus polarization Gaussian-type basis set was employed to expand the Kohn–Sham orbitals, and plane waves with a cutoff of 300 Ry were used to expand the charge density. The VASP code [45] was employed to calculate the optical properties of smaller models. The PBE functional [44] and projector augmented-wave (PAW) pseudopotentials [46] were used with an energy cutoff value of 500 eV. We considered a unit cell model for the hex-phase, and smaller supercell models for the cub-phase. The atomic structure of the amorphous phase was taken from our previous work [47]. The respective atomic models are shown in Fig. S1. Dense *k*-point meshes were used for the optical calculations, i.e., 16×16×1, 3×3×1 and 3×3×3 for the hex-, cub- and amorphous phase, respectively.

## 3. Results and discussion

### 3.1. In situ TEM characterization

Firstly, we carried out structural characterization experiments using a JEOL JEM-2100F TEM with a moderate accelerating voltage $V_a$ of 200 keV, and confirmed that the sample was in the cub-phase. Within a limited recording time of seconds, no obvious damage to the specimen was detected. Permanent structural changes become evident by focusing the electron beam to small exposure areas and extending the irradiation time to tens of minutes, as shown in Fig. 1. The focused electron beams have multiple effects, including specimen heating, knock-on collisions and radiolysis effects. It is generally agreed that the kinetic knock-on effects get stronger at higher accelerating voltages, while the heating effects become dominant at lower accelerating voltages [48]. For *in situ* electron beam irradiation experiments with properly tuned size of focus areas, including line scanning, operated under 200–300 keV, the formation and motion of swapped bilayer defects [49-51] and vacancy layers [52, 53], vacancy-disordering induced structural transition [54, 55], and even non-thermal amorphization [56] were observed in Ge-Sb-Te and $Sb_2Te_3$ alloys and superlattices.

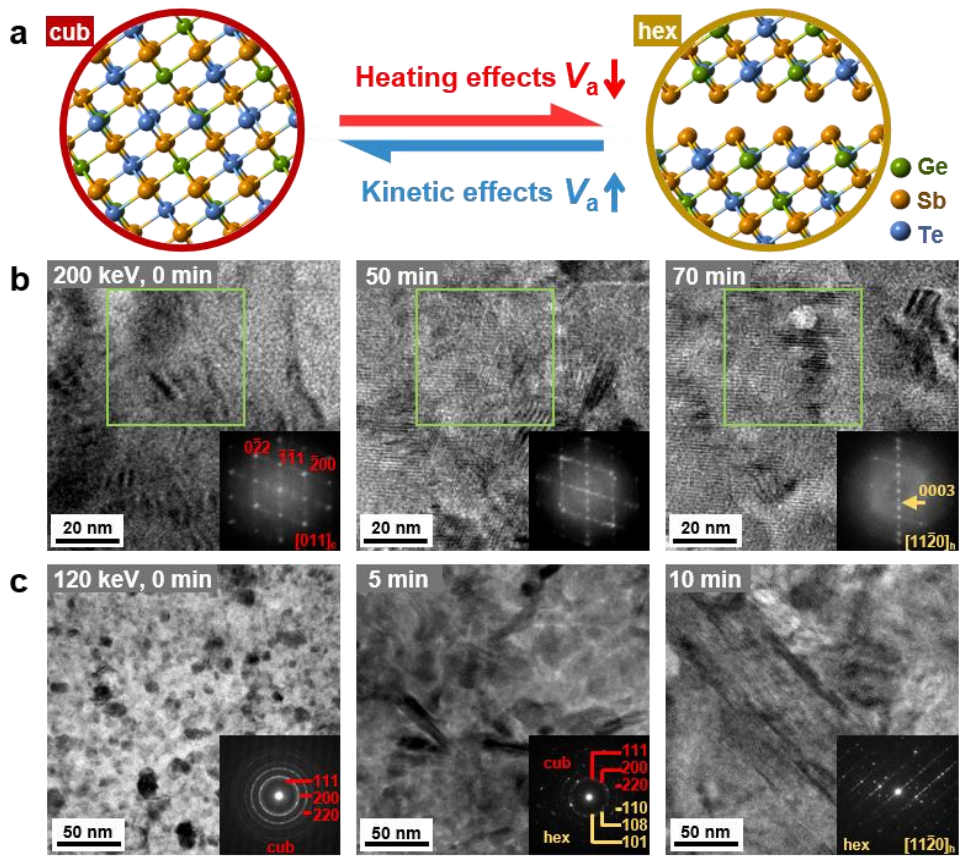

**Fig. 1. Structural phase transition in crystalline GST under electron beam irradiation.** (a) Sketch of the effects of electron beams in inducing reversible structural transition between the cub- and hex-phase of GST. (b) Bright-field TEM images and the FFT pattern corresponding to the area marked by the green box recorded at different irradiation time using the JEOL JEM-2100F TEM operated at 200 keV. (c) Bright-field TEM images and corresponding SAED patterns recorded at different irradiation time using the JEM-200CX TEM operated at 120 keV.

Using the JEOL JEM-2100F TEM operated at 200 keV, we managed to capture the cub-to-hex transition process within 70 min by tuning the beam radius for irradiating an area of slightly larger than 200 nm, corresponding to a beam intensity of ~1.0 × $10^{24}$ e $m^{-2}$ $s^{-1}$. We identified a cub-phase grain with the $[011]_c$ crystallographic orientation within the irradiated area, presented in Fig. 1b. In such view direction the cation-like and anion-like sublattices of the cubic rocksalt structure are well separated, and thus it is very suitable to observe vacancy ordering [18]. After extensive irradiation, planar stripes were observed in the bright-field images. The Fast Fourier Transform (FFT) analysis (insets) of the areas marked by green square shows a clear change in diffraction pattern from the cub- to hex-phase. To enhance the heating effects, we repeated the irradiation experiments on the same thin film by using a JEM-200CX TEM operated at 120 keV. As shown in Fig. 1c, the cub-to-hex transition took place faster and the majority of the irradiated area was transformed within ~10 min, as evidenced by the corresponding SAED patterns. The typical size of the irradiation-induced hex-phase grains is around a few hundreds of nanometers. Further irradiation will not induce visible changes inside the grains.

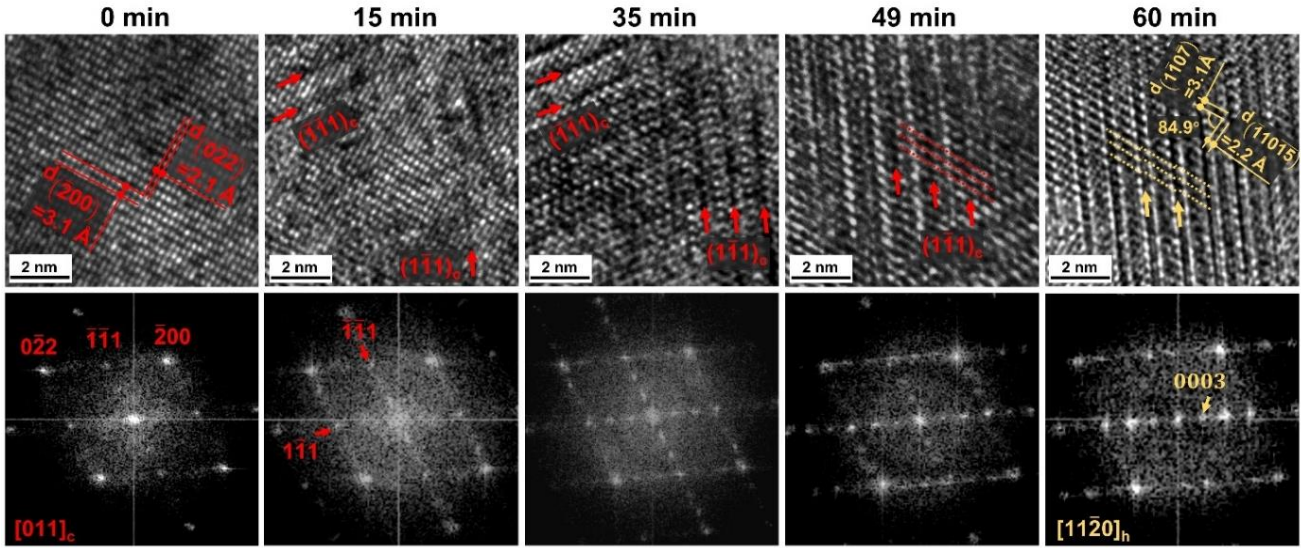

**Fig. 2. HRTEM images and corresponding FFT patterns of the local area shown in Fig. 1b.** Vacancy ordering is represented by appearance of the lattice periodicity of the hex-phase, as indicated by red and yellow arrows, respectively.

Next, we assess the detailed transition path of the structural transition process. Fig. 2 shows a local area within the green box presented in Fig. 1b with higher magnification. The initial phase looked homogenous, while after 15 min irradiation, some dark patches appeared in the $\{111\}_c$ planes, as marked by red arrows, which result in the reflection in the FFT pattern, $(\frac{\bar{1}}{3}\frac{1}{3}\frac{\bar{1}}{3})_c$ corresponding to $(0003)_h$, due to vacancy ordering. The intersection angle between the two dark patches is measured to ~71° in the HRTEM image. This observation of intersected dark patches with an angle of ~71° is consistent with the atomic-scale imaging experiments done on a $Ge_2Sb_2Te_5$ thin film annealed at 180 °C using HAADF-STEM [18]. As indicated by red arrows in the FFT pattern, these vacancy-ordered planes along the two different crystallographic planes are $(\bar{1}\bar{1}1)_c$ and $(1\bar{1}1)_c$ planes, both of which are equivalent to the $(111)_c$ planes. More dark patches along the two directions were observed after 35 min irradiation. Upon further irradiation to 49 min, vacancy ordering in the $(1\bar{1}1)_c$ planes became more evident but vanished in the $(\bar{1}\bar{1}1)_c$ planes. Later on, the atomic blocks were shifted locally when the concentration of vacancies reached a threshold value, inducing a change from the cub-phase stacking to the hexagonal one, which is consistent with our previous DFT prediction [36]. After 60 min irradiation, a hexagonal lattice with regular atomic blocks separated by structural gaps was observed. The corresponding FFT pattern shows the $[11\bar{2}0]_h$ view direction of the hexagonal lattice. Although it is not possible to determine the critical concentration of vacancies that induces the stacking change from the HRTEM experiments, the overall transition process is consistently observed, see Fig. S2. The process can be summarized into four stages, namely, (I) uncorrelated vacancy diffusion, (II) correlated vacancy ordering, (III) change in atomic stacking sequence and (IV) further vacancy and compositional ordering.

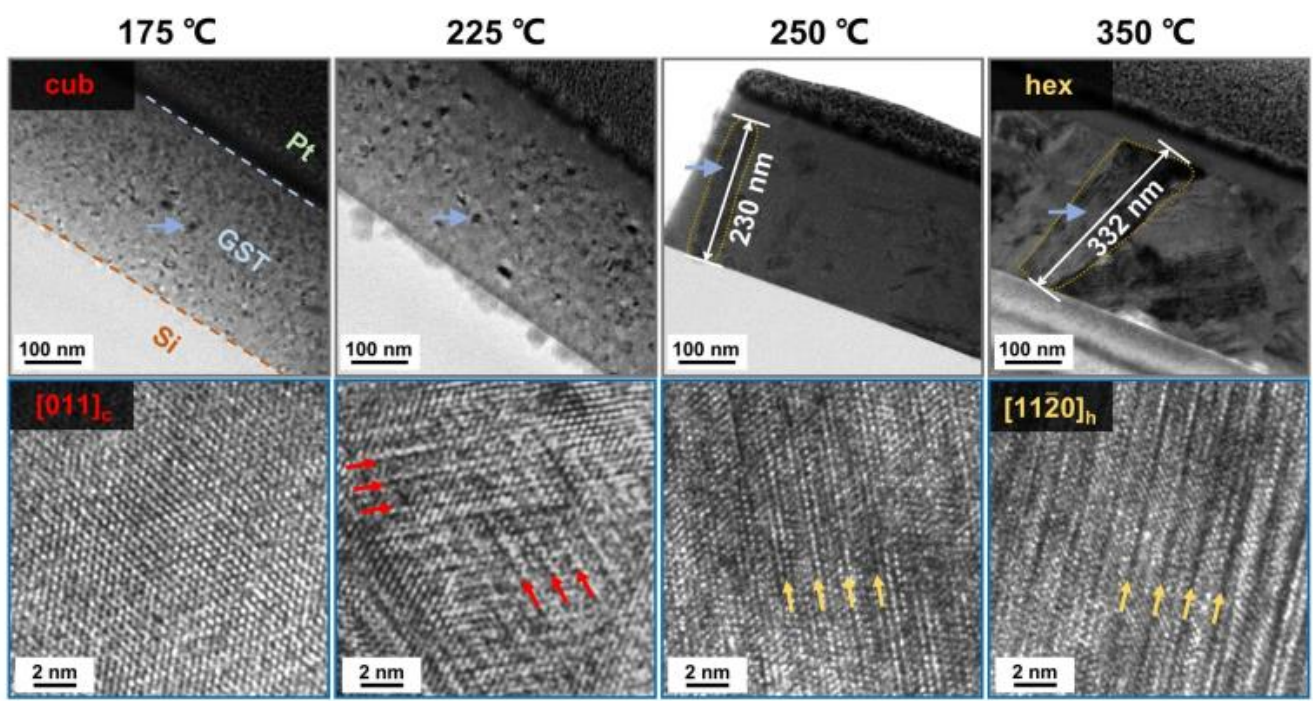

**Fig. 3. The *ex situ* TEM experiments upon thermal annealing in oven.** Vacancy ordering appears in the cub-phase grains in the thin film sample annealed at 225 °C (red arrows).

*3.2. Ex situ TEM characterization*

To confirm that this irradiation-induced phenomenon is mainly due to thermal effects, we also performed cross-sectional TEM characterization on the thin film samples annealed in oven. As shown in Fig. 3, the specimen annealed at 175 °C formed a standard cub-phase without any visible vacancy ordering, while the vacancy ordering process became evident in the specimen annealed at 225 °C. As indicated by the red arrows, the ordering of vacancies proceeded along the $\{111\}_c$ planes, consistent with the *in situ* observation. At this annealing temperature, both cub- and hex-phases coexist [35]. At higher annealing temperatures, large hex-phase grains with clear $GeSb_2Te_4$ blocks and structural gaps were observed. This set *ex situ* heating experiments was performed using thin films of ~450 nm, which shows consistent transition behavior as compared with the experiments carried out using thinner film of ~100 nm [35].

To obtain atomic-scale details on the transformed state, we carried out high-resolution HAADF-STEM imaging experiments on three sets of samples obtained by *in situ* irradiation experiments under different accelerating voltages and by *ex situ* heating experiments. As shown in Fig. S3, clear septuple-layer (SL) GeSb$_2$Te$_4$ blocks are revealed under *ex situ* heating (annealed at 300 °C) and under *in situ* irradiation using JEM-200 CX TEM, where heating effects dominated owing to the relatively lower accelerating voltage at 120 keV. However, quintuple-layer (QL) Sb$_2$Te$_3$ blocks were mostly found in the sampled irradiated in the JEOL JEM-2100F TEM operated at 200 keV, as shown in Fig. S3c. Due to the competing kinetic effect brought by the higher accelerating voltage, the surrounding matrix turned into amorphous phase [56] with Ge-richer composition. Inside the crystalline grain, the composition is deficient in Ge, accounting around 2.6 at% according to the EDX analysis. Therefore, the major structural fragment is SL, instead of QL. Nevertheless, the atomic details of vacancy ordering are the same for both Ge-Sb-Te alloys and the parent phase, Sb$_2$Te$_3$.

*3.3. Electronic and optical properties*

Next, we performed DFT calculations to gain further understanding of vacancy ordering in Ge-Sb-Te alloys. Following our previous work, we built three sets of orthorhombic supercell models as shown in Fig. 4a, where the vertical direction corresponds to the [111]$_c$ direction of the cub-phase [36]. Each model contains 144 Ge, 288 Sb, 576 Te and 144 vacancies. The first set of models contains randomly distributed vacancies occupying 25% of each cation-like lattice plane (denoted as cub-25%), and the second set includes 3 vacancy-rich (111)$_c$ planes containing 50% vacancies (denoted as cub-50%). In the third set of models, vacancies are distributed over the (111)$_c$ and its equivalent planes, and all the planes contain 50% vacancies (denoted as cub'-50%). For each configuration, five independent models with statistical distribution of vacancies, Ge and Sb atoms were generated and optimized by DFT relaxation calculations using CP2K. The obtained intersection angle between the two vacancy-rich planes is ~71°, in perfect agreement with the HRTEM observation. The average energy of the

cub-50% model is ~5.5 meV/atom lower than that of cub-25%, while cub'-50% is slightly lower in energy than cub-50% by ~1.8 meV/atom. This energy trend provides an explanation why vacancy ordering into both (111)$_c$ and its equivalent planes takes place in parallel at the early stage of the transition. Although from HRTEM images the specific concentration of vacancy cannot be determined for these vacancy-rich planes this uncorrelated vacancy ordering could proceed before the cub'-50% configuration is reached. In other words, it is feasible to build, e.g., cub'-30% or cub'-40% models. However, further ordering of vacancies must proceed via one set of {111}$_c$ planes, as it is no longer possible to increase the vacancy concentration in both sets of planes in parallel while keeping the correct stoichiometry. For instance, a hypothetical cub'-75% configuration in this orthorhombic supercell would request 216 vacancies, which is larger than the actual number of vacancies in our stoichiometric GeSb$_2$Te$_4$ model (144 vacancies).

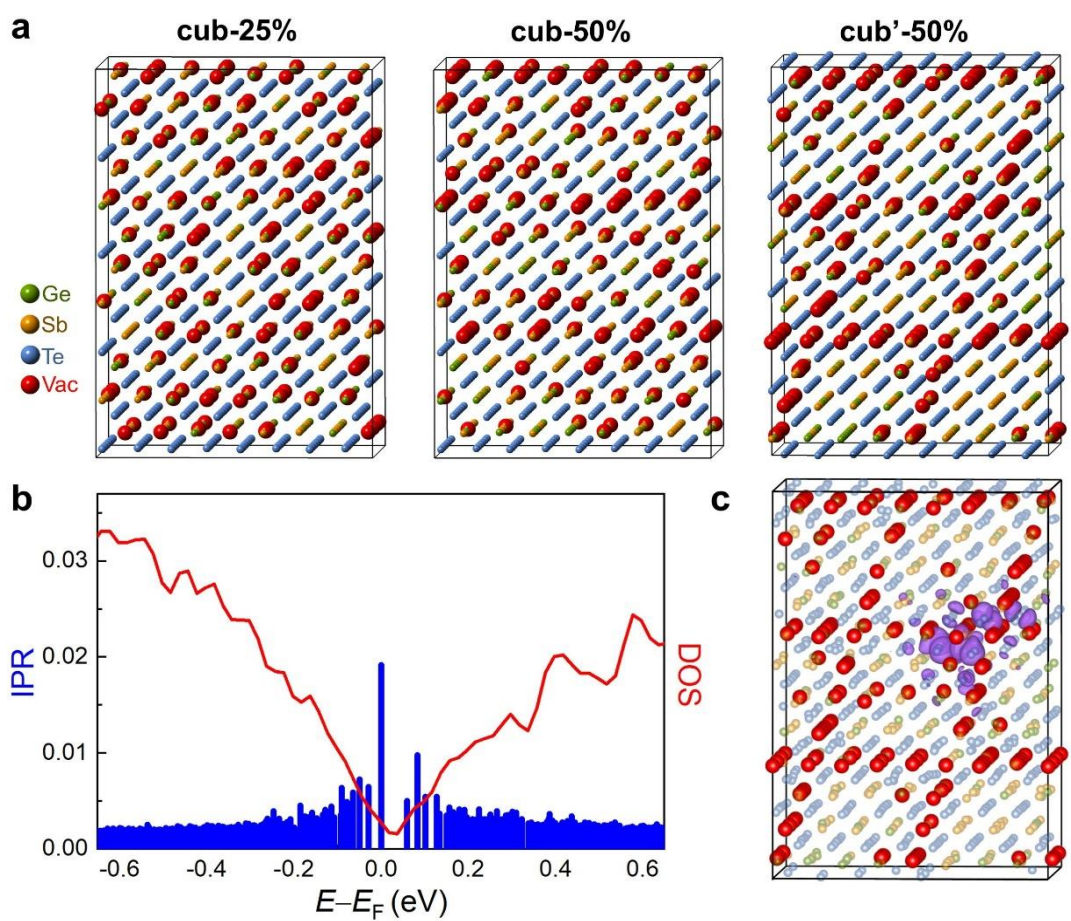

**Fig. 4. Atomic and electronic structure of one of the models with equivalent vacancy-rich planes.** (a) Atomic structures of representative cub-25%, cub-50% and cub'-50% models, respectively. The green, yellow, blue and red spheres represent Ge, Sb, Te and vacancies, respectively. (b) Calculated density of states (DOS) and inverse participation ratio (IPR) of the cub'-50% model. (c) Highest occupied molecular orbital (HOMO) of the cub'-50% model. The isosurface rendered in purple corresponds to an isovalue of 0.012 a.u.

We have shown that partial vacancy ordering at the stage of cub-50% is not sufficient to trigger the insulator-to-metal transition [36]. Here, we show that the cub'-50% configuration features Anderson localization of electrons, similar to that of the cub-50% case. Electron localization is consistently observed in all the five cub'-50% models as evidenced by DFT calculations. The calculated density of states (DOS) and inverse participation ratio (IPR) values of one such model is shown in Fig. 4b. The IPR is used to determine the degree of localization of a specific electronic state, i.e., the larger the IPR value, the more localized the state. Specifically, the IPR of a Kohn-Sham state $\Psi_\alpha$ is defined as $\Sigma_i |\Psi_{\alpha,i}|^4 / (\Sigma_i |\Psi_{\alpha,i}|^2)^2$, where the $\Psi_{\alpha,i}$'s indicate the expansion coefficients of $\Psi_\alpha$ with respect to the basis set. The DOS and IPR profiles shown in Fig. 4b reveal Anderson insulating behavior, in which the band tails contain localized states, i.e. states with a significantly larger IPR value, while the states away from the Fermi energy are delocalized. Fig. 4c shows that the highest occupied molecular orbital (HOMO) is localized around a vacancy-rich region, consistent with our previous studies [26, 36]. In short, the formation of equivalent vacancy-rich (111)$_c$ planes at the early stage does not yet suppress Anderson localization in the band tails in the cub-phase. We repeated the same set of calculations for $Sb_2Te_3$ (Fig. S4), and observed the same trend in total energy and electron localization upon vacancy ordering as in $GeSb_2Te_4$.

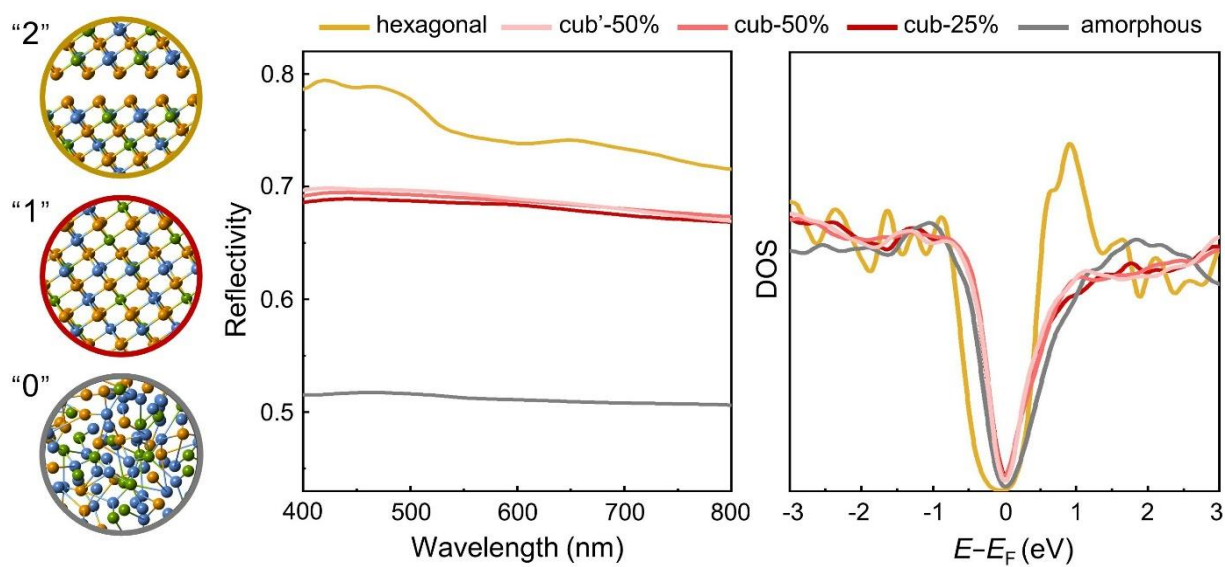

**Fig. 5. The calculated optical reflectivity and DOS for the hexagonal, cub-25%, cub-50%, cub'-50% and amorphous phase.** The atomic models shown in the left panel represent the hexagonal, cubic and amorphous phase, corresponding to the logic states "2", "1" and "0", respectively. The full atomic structures are shown in Fig. S1.

We also evaluated the impact of vacancy ordering on the optical reflectivity. For comparison, we also considered the optical response of the amor- and hex- phases. Smaller models were built for the calculation of the optical response using VASP. We calculated the frequency-dependent dielectric matrix within the independent-particle approximation without considering local field effects and many body effects, which have been shown not to affect the results significantly in similar compounds [57, 58]. Fig. 5 shows obvious contrast in the calculated optical reflectivity between the three phases. Conventional phase-change photonic devices mainly utilize the contrast between the amor-phase and the cub-phase, labeled as state "0" and "1". Given the similar DOS profile of the two phases, the significant contrast in reflectivity is attributed to the alignment (cub-phase) and misalignment (amor-phase) of the *p* orbitals of the Ge, Sb and Te atoms [59]. In particular, in cub-GST, the average number of *p* electrons is three per site, forming an extended network of highly aligned *p* orbitals, which leads to MVB [30-34]. Recently, the hex-GST alloyed with Nitrogen impurities was used as third logic state in non-volatile displays [60], in addition to the amor- and cub-phase. Our calculations indicate that the differences in reflectivity between the cub-25%, cub-50% and cub'-50% models are small. However, the hex-phase exhibits a much larger reflectivity, consistent with experimental observations [61]. As shown in the DOS profile, a notable peak is present at ~1 eV near the edge of the conduction band for the hex-phase, enhancing the optical inter-band excitations and, thus, the reflectivity.

*3.4. Multifold phase transitions*

At last, we provide a summary diagram of all the phase transition paths for Ge-Sb-Te alloys and Sb$_2$Te$_3$ in Fig. 6. Four phases are involved, namely the stable hexagonal phase, the metastable cubic and amorphous phases, and a transient liquid phase. Conventional phase-change memory makes use of Joule heating (thermal effect), induced by either electrical or optical pulses, to trigger the transition between the amorphous and cubic phase. Thermal effects directly drive the crystallization of the amorphous state into the cubic phase, while the reverse, "melt-quench" process has to go through the liquid state. By tuning the volume ratio between the amorphous and cubic phase using multiple weak pulses, multilevel states can be achieved via either partial amorphization (cubic → liquid → amorphous) or cumulative crystallization (amorphous → cubic), as shown in Fig. 6b. The former process defines multilevel states during programming in a more accurate manner, but suffers from the spontaneous structural relaxation (aging) of the amorphous phase [7]. The latter process allows a continuous and non-linear change in conductance and reflectivity, but is subjected to the high randomness associated with nucleation and grain growth into polycrystalline forms [2]. Both device operations are very important for multilevel storage and neuro-inspired and in-memory computing applications [62-64], and have been systematically improved via both materials optimization [65-67] and device engineering [68].

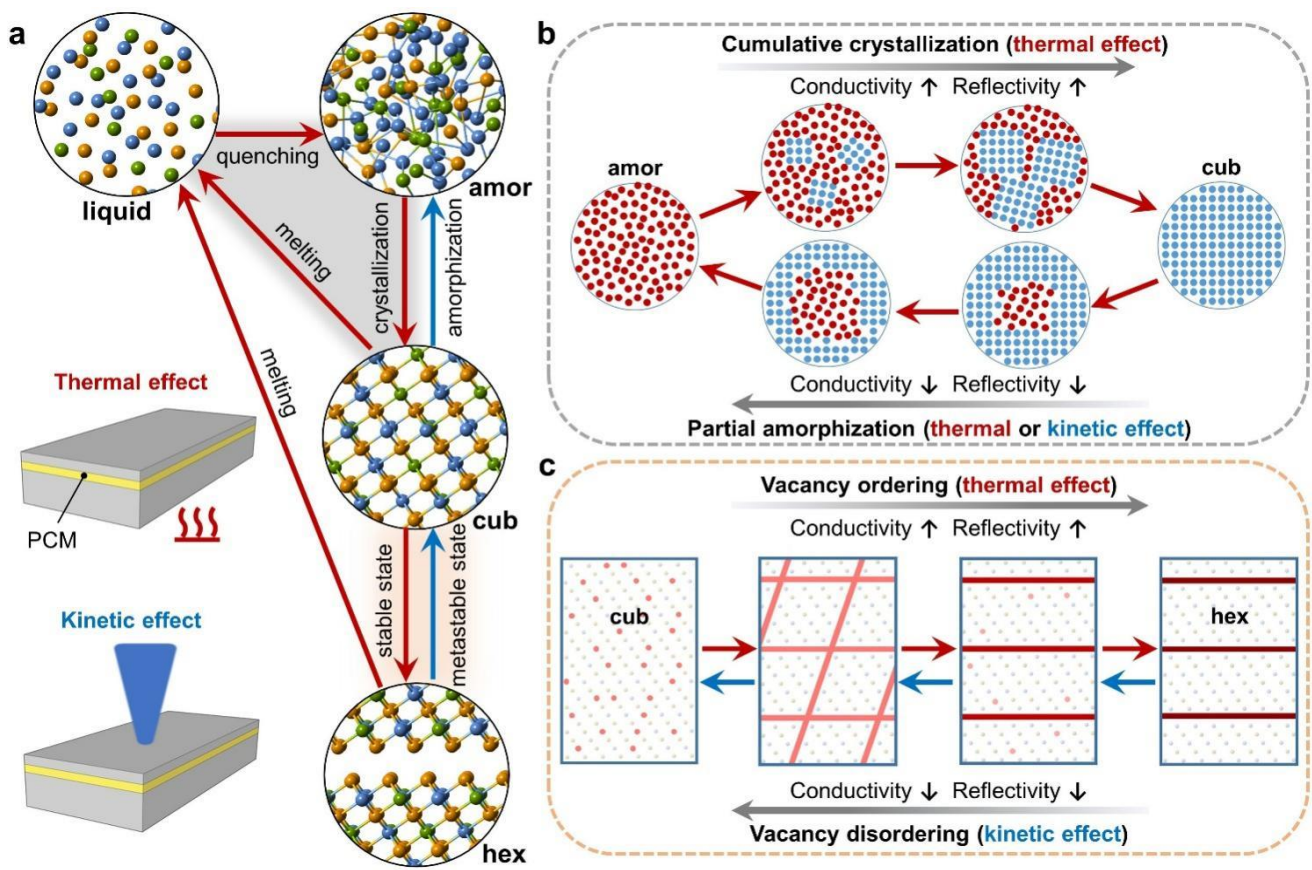

**Fig. 6. Phase transition paths for Ge-Sb-Te alloys and Sb₂Te₃.** (a) Summary of the phase transitions between the amorphous (amor), cubic (cub), hexagonal (hex) and liquid phase. The red and blue arrows indicate the paths triggered by thermal effects and kinetic effects, respectively. (b) shows the cumulative crystallization and partial amorphization to realize multilevel states in phase-change devices, involving the transition paths triggered by thermal effects marked in the grey shaded area in a. (c) Atomic details of vacancy ordering/disordering involved in the phase transition between the cubic and hexagonal phase. The transition paths are highlighted in the shaded orange area in a. The big red dots and the thick red lines represent vacancies and vacancy-rich planes, whereas smaller dots represent the atoms.

The thermal effect can also trigger the transition from the metastable cubic phase to the stable hexagonal phase (cubic → hexagonal), while the reverse transition by this effect is rather complex, i.e., hexagonal → liquid → amorphous → cubic. In contrast, a direct hexagonal → cubic transition can proceed via the knock-on kinetic effect induced by electron irradiation [54, 55] or ion bombardment [69]. Stronger kinetic effects can even trigger a direct amorphization process, bypassing the liquid phase (cubic → amorphous) [56]. The crystal-crystal transitions are mediated by either vacancy ordering or vacancy disordering, as outlined in Fig. 6c. The ordering of vacancies in the cubic phase first proceeds through an uncorrelated ordering in the equivalent vacancy-rich planes, and is followed by correlated ordering with only one set of dominant vacancy-rich planes. Further vacancy ordering drives block shifts to form the hexagonal stacking. This process can be reversed in the presence of sufficient knock-on collisions. The structural evolution towards the hexagonal phase largely extends the programming window for both electrical and optical devices, and does not require any change in programming volume. Given the absence of aging effects in the crystalline states, the vacancy ordering process could be useful for advanced memory and computing applications, although more thermal power and/or wider pulse widths are needed to form, as well as to melt down, these more stable crystalline states.

## 4. Conclusion

In summary, we have shown a progressive vacancy ordering mediated structural transition from cub- to hex-phase in GST and $Sb_2Te_3$ thin films under electron beam irradiation. The *in situ* HRTEM experiments present a comprehensive temporal and spatial view of the vacancy ordering process, and the DFT calculations unveil the energy origin and the associated changes in the electronic and optical properties along the transition path. Consistent with oven heating, the electron irradiation induced thermal effect triggers a progressive vacancy ordering and structural transformation, which proceeds via uncorrelated vacancy ordering within equivalent vacancy-rich planes, followed by correlated ordering with one set of vacancy-rich planes, and finally further ordering and block shifting. We also provide a comprehensive overview of all phase transition paths for Ge-Sb-Te alloys and $Sb_2Te_3$, induced by either thermal or kinetic agitation. The summary diagram can serve as a guide to tune the phase transition, and thereby, to tailor the properties of phase-change alloys for practical memory and computing applications.

**Acknowledgements**

The authors thank Danli Zhang and Ruihua Zhu for their technical support on TEM experiments. W.Z. thanks the support of National Natural Science Foundation of China (61774123). E.M. acknowledges the support of National Natural Science Foundation of China (52150710545). W.Z. and E.M. are grateful to XJTU for the support of their work at CAID. J.-J.W. and M.W. acknowledges financial support from Alexander von Humboldt Foundation. M.W. acknowledges funding from Deutsche Forschungsgemeinschaft within SFB 917 "Nanoswitches". W.Z. thanks the support of 111 Project 2.0 (BP2018008) and the International Joint Laboratory for Micro/Nano Manufacturing and Measurement Technologies of Xi'an Jiaotong University. The authors also acknowledge the computational resources provided by the HPC platform of Xi'an Jiaotong University and the Hefei Advanced Computing Center, and the National Supercomputing Center in Xi'an.
**Declaration of Competing Interest**

The authors declare no competing interests.

**References**

[1] M. Wuttig, N. Yamada, Phase-change materials for rewriteable data storage, Nat. Mater. 6 (2007) 824-832.

[2] W. Zhang, R. Mazzarello, M. Wuttig, et al., Designing crystallization in phase-change materials for universal memory and neuro-inspired computing, Nat. Rev. Mater. 4 (2019) 150-168.

[3] A. Sebastian, M. Le Gallo, R. Khaddam-Aljameh, et al., Memory devices and applications for in-memory computing, Nat. Nanotechnol. 15 (2020) 529–544.

[4] F. Rao, K. Ding, Y. Zhou, et al., Reducing the stochasticity of crystal nucleation to enable subnanosecond memory writing, Science 358 (6369) (2017) 1423-1427.

[5] J. Feldmann, N. Youngblood, C.D. Wright, et al., All-optical spiking neurosynaptic networks with self-learning capabilities, Nature 569 (2019) 208-214.

[6] Y. Zhang, J.B. Chou, J. Li, et al., Broadband transparent optical phase change materials for high-performance nonvolatile photonics, Nat. Commun. 10 (2019) 4279.

[7] W. Zhang, E. Ma, Unveiling the structural origin to control resistance drift in phase-change memory materials, Mater. Today 41 (2020) 156-176.

[8] J. Shen, S. Jia, N. Shi, et al., Elemental electrical switch enabling phase segregation–free operation, Science 374 (6573) (2021) 1390–1394.

[9] X.-D. Wang, W. Zhang, E. Ma, Monatomic phase-change switch, Sci. Bull. 67 (9) (2022) 888-890.

[10] M. Xu, X. Mai, J. Lin, et al., Recent advances on neuromorphic devices based on chalcogenide phase-change materials, Adv. Funct. Mater. 30 (50) (2020) 2003419.

[11] N. Yamada, E. Ohno, K. Nishiuchi, et al., Rapid-phase transitions of GeTe-$Sb_2Te_3$ pseudobinary amorphous thin films for an optical disk memory, J. Appl. Phys. 69 (5) (1991) 2849-2856.

[12] T. Matsunaga, N. Yamada, Structural investigation of $GeSb_2Te_4$: A high-speed phase-change material, Phys.

## Supplementary Material

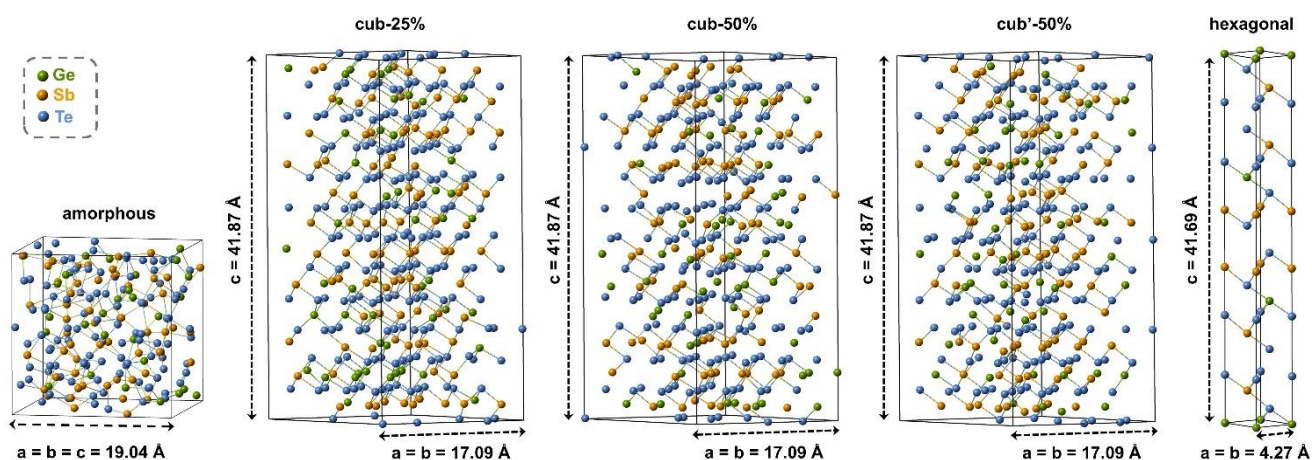

**Figure S1.** DFT relaxed structures for the amorphous, cub-25%, cub-50%, cub'-50% and hexagonal phase. Parts of the atomic structures are shown in Figure 5. The amorphous model contains 189 atoms. The cubic supercell models contain 336 atoms. The hexagonal unit cell model contains 21 atoms.

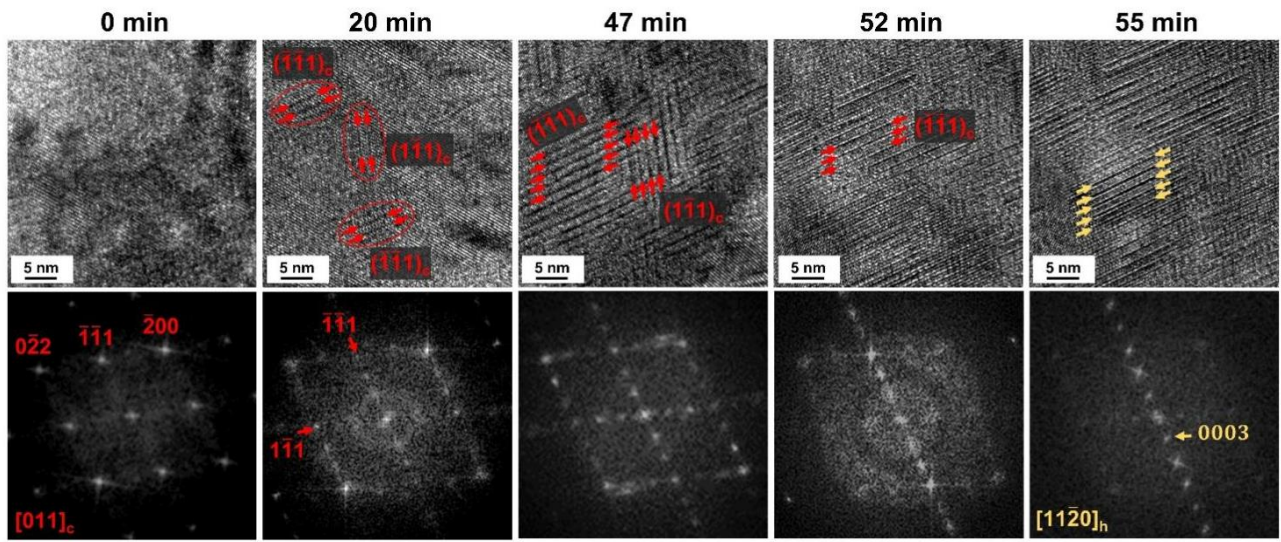

**Figure S2.** *In situ* observation of vacancy ordering in another local area extracted from Figure 1b. Vacancy ordering appears in the {111}$_c$ planes of the cub-phase (red arrows) and in hex-phase lattices indicated by yellow arrows.

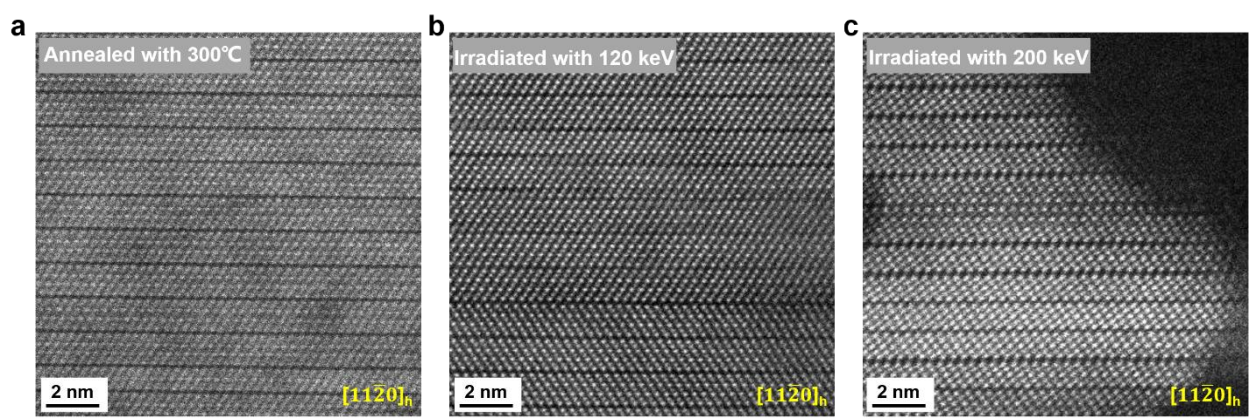

**Figure S3.** The HAADF images on (a) the thermally annealed sample at 300 °C, (b) the irradiated sample performed on JEM-200 CX TEM operated at 120 keV, and (c) the irradiated sample performed on JEOL JEM-2100F TEM operated at 200 keV. The former two samples show SL blocks, while the latter shows QL blocks. The dark regions in (c) are in the amorphous phase with Ge-richer composition.

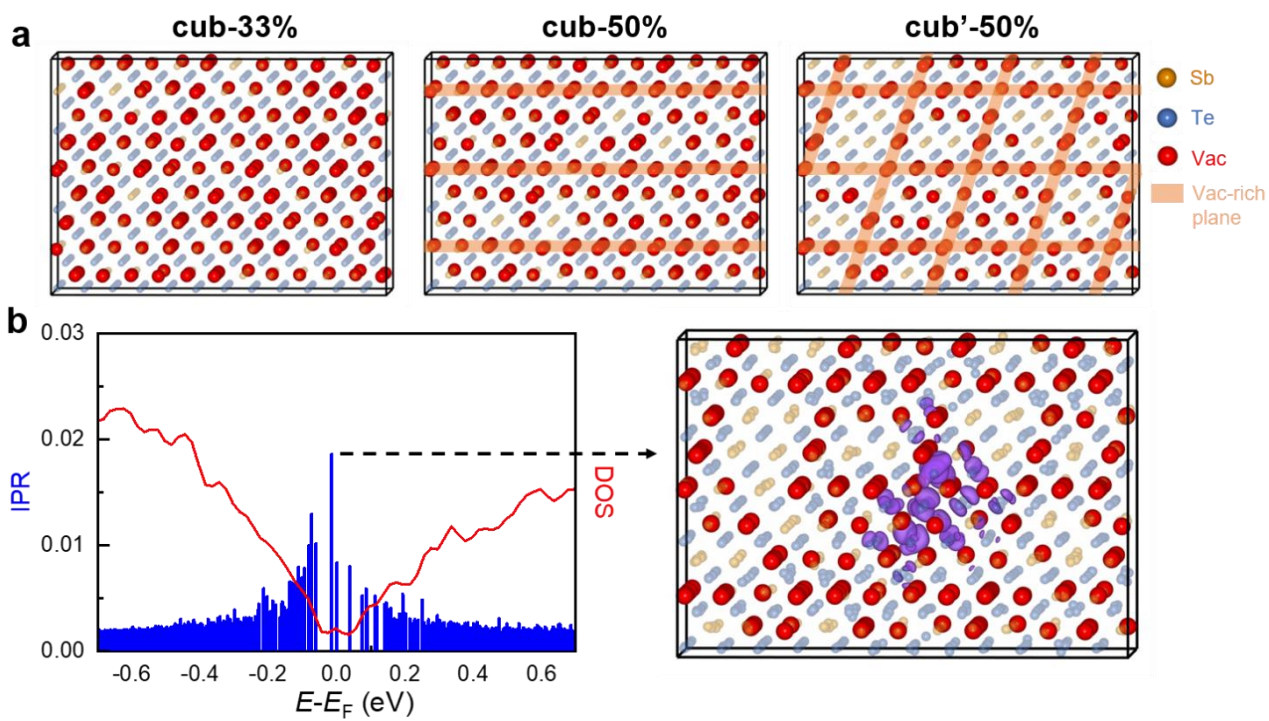

**Figure S4.** Atomic and electronic structure of Sb$_2$Te$_3$. (a) Atomic structures of cub-33%, cub-50% and cub'-50%, respectively. Each model contains 1080 atoms. The yellow, blue and red spheres represent Sb, Te and vacancy, respectively. The vacancy-rich planes are highlighted in orange. Five independent models were considered. The average total energy values show that cub-50% is ~5.6 meV/atom lower than cub-33%, while cub'-50% is slightly lower in energy than cub-50% by ~1.5 meV/atom. (b) The calculated DOS, IPR and charge isosurface of an occupied state (indicated by the dashed arrow) of the cub'-50% model. The isosurface rendered in purple corresponds to an isovalue of 0.012 a.u.